\documentclass[journal]{IEEEtran}

\usepackage{amssymb}
\usepackage{caption2}
\usepackage{amsmath}
\usepackage{multirow}
\usepackage{amsthm}
\usepackage[ruled,vlined,commentsnumbered]{algorithm2e}
\newtheorem{theorem}{Theorem}[section]
\newtheorem{definition}[theorem]{Definition}
\newtheorem{lemma}[theorem]{Lemma}

\newtheorem{corollary}[theorem]{Corollary}

\begin{document}

\title{Codes with the Identifiable Parent Property for Multimedia Fingerprinting}

\author{Minquan~Cheng,~Hung-Lin Fu, ~Jing Jiang, ~Yuan-Hsun Lo and Ying Miao
\thanks{The research of Cheng was supported by NSFC (No.11301098),
Guangxi Natural Science Foundations (No.2013GXNSFCA019001), and the Scientific Research Foundation for the Returned Overseas Chinese Scholars, State Education Ministry. The research of Fu and Lo was supported by NSC 100-2115-M-009-005-MY3. The research of Miao was supported by JSPS Grant-in-Aid for Scientific Research (C) under Grant No.~24540111.
}
\thanks{M. Cheng is with Department of Mathematical Sciences, Guangxi Normal University, Guilin 541004, P. R. China.
E-mail: chengqinshi@hotmail.com.}
\thanks{H-L Fu is with Department of Applied Mathematics, National Chiao Tung University, Hsinchu 300, Taiwan.
E-mail: hlfu@math.nctu.edu.tw.}
\thanks{J. Jiang and Y. Miao are with Department of Social Systems and Management,
Graduate School of Systems and Information Engineering, University of Tsukuba, Tsukuba 305-8573, Japan.
E-mails: jjiang2008@hotmail.com, miao@sk.tsukuba.ac.jp.}
\thanks{Y-H Lo is with Department of Mathematics,  National Taiwan Normal University, Taipei $116$, Taiwan.
E-mail: yhlo0830@gmail.com.}
\thanks{Copyright (c) 2014 IEEE. Personal use of this material is permitted. However, permission to use this material for any other purposes must be obtained from the IEEE by sending a request to pubs-permissions@ieee.org.}
}
\maketitle

\begin{abstract}
Let ${\cal C}$ be a $q$-ary code of length $n$ and size $M$, and
${\cal C}(i) = \{{\bf c}(i) \ | \ {\bf c}=({\bf c}(1), {\bf c}(2), \ldots,  {\bf c}(n))^{T} \in {\cal C}\}$
be the set of $i$th coordinates of ${\cal C}$.
The descendant code of a sub-code ${\cal C}^{'} \subseteq {\cal C}$ is defined to be
${\cal C}^{'}(1) \times {\cal C}^{'}(2) \times \cdots \times {\cal C}^{'}(n)$.
In this paper, we introduce a multimedia analogue of codes with the identifiable parent property (IPP), called multimedia IPP codes or
$t$-MIPPC$(n, M, q)$, so that
given the descendant code of any sub-code ${\cal C}^{'}$ of a multimedia $t$-IPP code ${\cal C}$,
one can always identify, as IPP codes do in the generic digital scenario, at least one codeword in ${\cal C}^{'}$.
We first derive a general upper bound on the size $M$ of a multimedia $t$-IPP code,
and then investigate multimedia $3$-IPP codes in more detail.
We characterize a multimedia $3$-IPP code of length $2$ in terms of a bipartite graph and a generalized packing, respectively.
By means of these combinatorial characterizations,
we further derive a tight upper bound on the size of a multimedia $3$-IPP code of length $2$,
and construct several infinite families of (asymptotically) optimal  multimedia $3$-IPP codes of length $2$.
\end{abstract}


\section{Introduction}
\label{intro}

Let $n \ge 2$, $M$ and $q \ge 2$ be positive integers, and $Q$ an alphabet with $|Q|=q$.
In this paper, we consider a code ${\cal C}$ of length $n$ over $Q$, that is,
a set ${\cal C} = \{{\bf c}_1,{\bf c}_2,\ldots, {\bf c}_M\} \subseteq Q^n$.
Each ${\bf c}_i$ in such an $(n,M,q)$ code is called a codeword.
Without loss of generality, we may assume $Q=\{0,1,\ldots,q-1\}$.
Given an $(n,M,q)$ code, its incidence matrix is the $n \times M$ matrix on $Q$
in which the columns are the $M$ codewords in ${\cal C}$.
Often, we make no difference between an $(n,M,q)$ code and its incidence matrix.

For any code ${\cal C} \subseteq Q^n$, we define the set of $i$th coordinates of ${\cal C}$ as
\[ {\cal C}(i) =\{{\bf c}(i) \in Q \ | \ {\bf c}=({\bf c}(1), {\bf c}(2), \ldots, {\bf c}(n))^{T} \in {\cal C}\}\]
for any $1 \le i \le n$.
For any sub-code ${\cal C}^{'} \subseteq {\cal C}$, we define the descendant code of ${\cal C}^{'}$ as
\begin{eqnarray*}
 {\sf desc}({\cal C}^{'}) = \{({\bf x}(1), {\bf x}(2), \ldots, {\bf x}(n) )^{T}  \in Q^n \ |\ \ \ \ \ \ \ \ \ \ \ \ \ \ \ \ \ \   \\ \ \ \ \ \ \ \ \ \ \ \ \ \ \ \ \ \ \ \ \ \ \ \ \ \ \ \ \ \ \ \ \ \ \ \ \  {\bf x}(i) \in {\cal C}^{'}(i), 1 \le i \le n\},\end{eqnarray*}
that is,
\[ {\sf desc}({\cal C}^{'})={\cal C}^{'}(1) \times {\cal C}^{'}(2) \times \cdots \times  {\cal C}^{'}(n).\]
Any codeword in ${\cal C}^{'}$ is a parent of all the words in ${\sf desc}({\cal C}^{'})$.

\begin{definition}
\label{MIPP}
Let $\cal C$ be an $(n,M,q)$ code, and for any $S \subseteq {\cal C}(1) \times {\cal C}(2) \times \cdots \times {\cal C}(n)$,
define the set of parent sets of $S$ as
\[ {\cal P}_{t}(S) = \{{\cal C}^{'} \subseteq {\cal C} \ | \ |{\cal C}^{'} | \leq t, S = {\sf desc}({\cal C}^{'})\}. \]
We say that $\cal C$ is a code with the identifiable parent property (IPP) for multimedia fingerprinting, or a multimedia IPP code, denoted $t$-MIPPC$(n,M,q)$, if
\[ \bigcap_{{\cal C}^{'} \in {\cal P}_{t}(S)}{\cal C}^{'} \neq \emptyset \]
is satisfied for any $S \subseteq {\cal C}(1) \times {\cal C}(2) \times \cdots \times {\cal C}(n)$ with ${\cal P}_{t}(S) \neq \emptyset$.
\end{definition}

Intuitively, ${\cal P}_t(S)$ consists of all the sub-codes of ${\cal C}$ with size at most $t$
that could have produced all the words in $S$, and an $(n,M,q)$ code ${\cal C}$ is a $t$-MIPPC$(n,M,q)$
if the following condition is satisfied:
even if there are distinct sub-codes of ${\cal C}$, each of size at most $t$, could produce the same set $S$ of words,
we can track down at least one parent of $S$ which is contained in each parent set of $S$.
In fact, any codeword in $\bigcap_{{\cal C}^{'} \in {\cal P}_{t}(S)}{\cal C}^{'}$ is a parent of $S$.

Multimedia IPP codes are a variation of IPP codes and a generalization of separable codes,
both were introduced for the purpose of protecting copyrighted digital contents.
The notion of an IPP code was first introduced in a special case in \cite{HLLT},
investigated in full generality in \cite{BCEKZ, BK, Bl, SSW, TM},  and surveyed in \cite{B}.
The notion of a separable code was introduced in \cite{CM} and investigated in detail in \cite{CJM, GG}.
In Definition \ref{MIPP}, if $S$ is set to be a singleton set $\{{\bf d}\}$, and the set of parent sets be modified as
\[ {\cal P}_{t}(S) = \{{\cal C}^{'} \subseteq {\cal C} \ | \ |{\cal C}^{'} | \leq t, {\bf d} \in {\sf desc}({\cal C}^{'})\}, \]
then we obtain a $t$-IPP code,
while if we require that $|{\cal P}_{t}(S)| = 1$ for any
$S \subseteq {\cal C}(1) \times {\cal C}(2) \times \cdots \times {\cal C}(n)$ with ${\cal P}_t(S) \ne \emptyset$,
then we obtain a $\overline{t}$-separable code.

Binary $\overline{t}$-separable codes are used in multimedia fingerprinting to capture up to $t$ malicious authorized users
holding the same multimedia content but with different codewords (i.e., fingerprints),
who have jointly produced a pirate copy of the copyrighted content from their authorized copies (see, for example, \cite{CM}).
However, in most cases, the number of codewords in a binary $\overline{t}$-separable code is too small to be of practical use.
Meanwhile, guaranteeing exact identification of at least one member of the coalition of size at most $t$
would bring enough pressure to bear on authorized users to give up their attempts at collusion.
Using the tracing algorithm {\tt MIPPCTraceAlg$(S)$} described in Section \ref{pre},
we know that by means of a binary multimedia IPP code, we can capture a set $S \subseteq {\cal C}(1) \times \cdots \times {\cal C}(n)$ in the multimedia scenario instead of an element ${\bf d} \in S$ in the generic digital scenario,
and although binary multimedia $t$-IPP codes can not identify all malicious users as binary $\overline{t}$-separable codes do when the size of the coalition is at most $t$,
they can identify, as IPP codes do in the generic digital scenario \cite{BBK, HLLT}, at least one such malicious authorized user,
thereby helping stop the proliferation of the fraudulent content in digital marketplace.

Therefore, we can say that in some sense, the significance of multimedia $t$-IPP codes relies on their maximum sizes.
For $t=2$, we will show in Lemma \ref{rela112} that a $t$-MIPPC$(n,M,q)$ is in fact a $\overline{t}$-SC$(n,M,q)$,
so they have the same maximum size.
For $t >2$, the maximum size of a $\overline{t}$-SC$(n,M,q)$ is $O(q^{\lceil n/(t-1) \rceil})$ (see \cite{CJM}),
while the maximum size of a $t$-MIPPC$(n,M,q)$ will be shown in Section \ref{tESC} to be $O(q^{(t+1)n/(2t)})$,
except for the case that $t$ is even and $n$ is odd, where the value is $O(q^{((t+1)n+1)/(2t)})$.
This is a significant improvement on the number of codewords, which makes the notion of multimedia IPP codes useful.

\begin{lemma}
\label{rela112}
Let $\cal C$ be an $(n,M,q)$ code.
Then $\cal C$ is a $2$-MIPPC$(n,M,q)$ if and only if it is a $\overline{2}$-SC$(n,M,q)$.
\end{lemma}
\begin{IEEEproof} It is clear that a $\overline{t}$-SC$(n,M,q)$ is necessary a $t$-MIPPC$(n,M,q)$.
We only need to consider its necessity.
Assume that $\cal C$ is a $2$-MIPPC$(n,M,q)$ such that ${\cal C}_1, {\cal C}_2 \subseteq \cal C$,
$|{\cal C}_1| \leq 2$, $|{\cal C}_2| \leq 2$, ${\cal C}_1 \neq {\cal C}_2$, and ${\sf desc}({\cal C}_1) = {\sf desc}({\cal C}_2)$.
Then ${\cal C}_1 \bigcap {\cal C}_2 \neq \emptyset$.
Let ${\bf a} \in {\cal C}_1 \bigcap {\cal C}_2$. There are two cases to be considered.
\begin{itemize}
\item[(1)] ${\cal C}_1 = \{{\bf a}\}$, ${\cal C}_2 = \{{\bf a, b}\}$: Since ${\sf desc}({\cal C}_1) = {\sf desc}({\cal C}_2)$,
we have ${\bf a} = {\bf b}$, which implies ${\cal C}_1 = {\cal C}_2$.
\item[(2)] ${\cal C}_1 = \{{\bf a, b}\}$, ${\cal C}_2 = \{{\bf a, c}\}$:
Let ${\bf a} = ({\bf a}(1), \ldots, {\bf a}(n))^T$, ${\bf b} = ({\bf b}(1), \ldots, {\bf b}(n))^T$ and ${\bf c} = ({\bf c}(1), \ldots, {\bf c}(n))^T$.
Since ${\sf desc}({\cal C}_1) = {\sf desc}({\cal C}_2)$, we have $\{{\bf a}(i), {\bf b}(i)\} = \{{\bf a}(i), {\bf c}(i)\}$ for any $1 \leq i \leq n$.
Now, if $ {\bf b}(i) = {\bf a}(i)$, then $ {\bf c}(i) = {\bf b}(i)$.
On the other hand, if ${\bf b}(i) \neq {\bf a}(i)$, then ${\bf c}(i) = {\bf b}(i)$ since $\{ {\bf a}(i), {\bf b}(i)\} = \{ {\bf a}(i), {\bf c}(i)\}$.
Hence, ${\bf c}(i) = {\bf b}(i)$ holds for any $1 \leq i \leq n$.
This implies ${\bf b} = {\bf c}$ and thus ${\cal C}_1 = {\cal C}_2$.
\end{itemize}
So for any distinct ${\cal C}_1, {\cal C}_2 \subseteq \cal C$ such that $|{\cal C}_1| \leq 2$, $|{\cal C}_2| \leq 2$,
it always holds that ${\sf desc}({\cal C}_1) \neq  {\sf desc}({\cal C}_2)$.
This means that ${\cal C}$ is a $\overline{2}$-SC$(n,M,q)$.
\end{IEEEproof}

In subsequent sections, we investigate the maximum size of a $t$-MIPPC$(n,M,q)$ and
also the constructions of (asymptotically) optimal $t$-MIPPC$(n,M,q)$s.
Let $M(t,n,q)$ denote the maximum size of a $t$-MIPPC$(n, M, q)$.
A $t$-MIPPC$(n,M,q)$ is said to be optimal if $M = M(t,n,q)$, and
asymptotically optimal if $\lim_{q \rightarrow \infty} \frac{M}{M(t,n,q)}=1$.
In Section \ref{pre}, we briefly review some terminologies,
describe a tracing algorithm based on binary multimedia IPP codes,
and show a construction for
binary multimedia IPP codes from $q$-ary multimedia IPP codes.
In Section \ref{tESC}, we derive a general upper bound on $M(t,n,q)$.
Then in Section \ref{3ESC}, we characterize $3$-MIPPC$(2,M,q)$s in terms of bipartite graphs and generalized packings, respectively.
The first graph theoretic characterization gives a tight upper bound on $M(3,2,q)$.
The second design theoretic characterization results in a construction of $3$-MIPPC$(2,M,q)$s,
in which some are optimal and some are asymptotically optimal.

\section{Preliminaries } %
\label{pre}                                                                  %

In this section, we give a brief review on some basic terminologies.
The interested reader is referred to \cite{CM,LTWWZ} for more detailed information.
We also describe a tracing algorithm based on binary multimedia IPP codes, and a construction for
binary multimedia IPP codes from $q$-ary multimedia IPP codes.

In general, collusion-resistant fingerprinting requires the
design of fingerprints that can survive collusion attacks to
trace and identify colluders, as well as robust embedding of
fingerprints into multimedia host signals.
One of the widely employed robust embedding techniques is spread-spectrum additive embedding,
which can survive collusion attacks to trace and identify colluders.
In spread-spectrum embedding, a watermark signal, often represented by
a linear combination of noise-like orthonormal basis signals, is added to the host signal.
Let ${\bf x}$ be the host multimedia signal, $\{ {\bf u}_i \ | \ 1 \leq i \leq n\}$ be an orthonormal basis of noise-like signals,
and $\{{\bf w}_j = ({\bf w}_j(1), {\bf w}_j(2), \ldots, {\bf w}_j(n)) = \sum_{i=1}^{n}b_{ij}{\bf u}_i\ | \ 1 \leq j \leq M\}$, $b_{ij} \in \{0,1\}$,
be a family of scaled watermarks to achieve the imperceptibility as well as to control the energy of the embedded watermark.
Each authorized user $U_j$, $1 \leq j \leq M$, who has purchased the rights to access ${\bf x}$,
is then assigned with a watermarked version of the content ${\bf y}_j = {\bf x} + {\bf w}_j$.
The fingerprint ${\bf w}_j$  assigned to $U_j$ can be represented uniquely by a vector (called codeword)
${\bf b}_j = (b_{1j}, b_{2j}, \ldots, b_{nj})^{T} \in \{0,1\}^{n}$ because of the linear independence
of the basis $\{ {\bf u}_i \ | \ 1 \leq i \leq n\}$.
Since distinct codes correspond to distinct fingerprinting strategies, we would like to strategically design a code
to accurately identify the contributing fingerprints involved in collusion attacks.

When $t$ authorized users, say $U_{j_1}, U_{j_2}, \ldots, U_{j_t}$,
who have the same host content  but distinct fingerprints come together,
we assume that they have no way of manipulating the individual orthonormal signals,
that is, the underlying codeword needs to be taken and proceeded as a single entity,
but they can carry on a linear collusion attack to generate a pirate copy from their $t$ fingerprinted contents,
so that the venture traced by the pirate copy can be attenuated.
For fingerprinting through additive embedding, this is done by linearly combining the $t$ fingerprinted contents
$\sum_{l=1}^{t}\lambda_{j_l}{\bf y}_{j_l}$, where the weights $\{{\lambda}_{j_l} \ | \ 1 \leq l \leq t\}$
satisfy the condition $\sum_{l=1}^{t}\lambda_{j_l} = 1$ to maintain the average intensity of the original multimedia signal.
In such a collusion attack, the energy of each of the watermarks ${\bf w}_{j_l}$ is reduced by a factor of $\lambda_{j_l}^{2}$,
therefore, the trace of $U_{j_l}$'s fingerprint becomes weaker and thus $U_{j_l}$ is less likely to be caught by the detector.
In fact, since normally no colluder is willing to take more of a risk than any other colluder,
the fingerprinted signals are typically averaged with an equal weight for each user.
Averaging attack choosing ${\lambda}_{j_l} = 1/t$, $1 \leq l \leq t$, is the most fair choice
for each colluder to avoid detection, as claimed in \cite{LTWWZ,TWWL}.
This attack also makes the pirate copy have better perceptional quality.

Based on the averaging attack model, the observed content ${\bf y}$ after collusion is
\[ {\bf y} = \frac{1}{t}\sum\limits_{l=1}^{t}{\bf y}_{j_l} = \frac{1}{t}\sum\limits_{l=1}^{t}{\bf w}_{j_l} + {\bf x}=
 \sum\limits_{l=1}^{t}\sum\limits_{i=1}^{n}\frac{b_{ij_l}}{t}{\bf u}_{i} + {\bf x}.\]

Due to the orthogonality of the orthonormal basis $\{{\bf u}_i \ | \ 1 \leq i \leq n\}$, in colluder detection phase,
we only need to compute the correlation vector
${\bf T} = ( {\bf T}(1), {\bf T}(2), \ldots, {\bf T}(n))$,
where ${\bf T}(i) = \langle {\bf y}-{\bf x}, {\bf u}_{i}\rangle$, $1 \leq i \leq n$,
and $\langle {\bf y}-{\bf x}, {\bf u}_{i}\rangle$ is the inner product of ${\bf y}-{\bf x}$ and  ${\bf u}_i$.

For any set of colluders holding codewords ${\cal C}_0 \subseteq {\cal C}$ and any index $1 \leq i \leq n$,
their detection statistics ${\bf T}(i)$ possesses the whole information
on ${\cal C}_0(i)$; namely, we have ${\bf T}(i)=1$ if and only if ${\cal C}_0(i)=\{1\}$,
${\bf T}(i)=0$ if and only if ${\cal C}_0(i)=\{0\}$, and
$0 < {\bf T}(i) < 1$ if and only if ${\cal C}_0(i)=\{0,1\}$.

Now we describe a tracing algorithm based on a binary multimedia IPP code.
The following theorem shows that binary multimedia $t$-IPP codes can be used to identify at least one colluder in the averaging attack.

\begin{theorem}
\label{ELACCAlg}
Under the assumption that the number of colluders in the averaging attack is at most $t$,
any $t$-MIPPC$(n, M, 2)$ can be used to identify at least one colluder  with computational complexity $O(nM^{t})$
by applying Algorithm \ref{al93} described below.
\end{theorem}
\begin{IEEEproof} Let $\cal C$ be the $t$-MIPPC$(n, M, 2)$,
and $S \subseteq {\cal C}(1) \times \cdots \times {\cal C}(n)$
be the captured descendant code derived from the detection statistics ${\bf T}$.
Then by applying the following tracing algorithm, Algorithm \ref{al93}, we can identify at least one colluder.
\begin{algorithm}[h]
\caption{ {\tt MIPPCTraceAlg}$({S})$}
\label{al93}
Given $S$\;
Find\ ${\cal P}_{t}(S) = \{{\cal C}^{'} \subseteq {\cal C} \ | \ |{\cal C}^{'} | \leq t, S = {\sf desc}({\cal C}^{'})\}$\;

Compute ${\cal C}_0 =\bigcap\limits_{{\cal C}^{'} \in {\cal P}_{t}(S)}{\cal C}^{'}$\;\vskip 0.2cm
\eIf{$|{\cal C}_0| \leq t$ }
{{\bf output} ${\cal C}_0$ as the set of colluders\;}{{\bf output} ``the set of colluders has size at least $t+1$"\;}
\end{algorithm}

The computational complexity is obvious.
We need only to show that any user $u$ assigned with a codeword ${\bf c} \in {\cal C}_0$ is a colluder.
Since $S$ is the captured descendant code derived from the detection statistics ${\bf T}$,
it is clear that ${\cal P}_{t}(S) \neq \emptyset$. Therefore,
$${\cal C}_0 =\bigcap\limits_{{\cal C}^{'} \in {\cal P}_{t}(S)}{\cal C}^{'} \neq \emptyset$$
by the definition of a multimedia $t$-IPP code.
Assume that $u$ is not a colluder.
Then for any ${\cal C}^{'} \in {\cal P}_{t}(S)$,
we have ${\cal C}^{'} \setminus \{{\bf c}\} \in {\cal P}_{t}(S)$,
which implies ${\bf c} \notin {\cal C}_0$, a contradiction.
\end{IEEEproof}

The following theorem is a simple composition construction for binary multimedia $t$-IPP codes
from $q$-ary multimedia $t$-IPP codes.

\begin{lemma}
\label{reconstru}
If there exists a $t$-MIPPC$(n,M,q)$, then there exists a $t$-MIPPC$(nq,M,2)$.
\end{lemma}
\begin{IEEEproof}
Let ${\cal C} = \{{\bf c}_1, {\bf c}_2, \ldots, {\bf c}_M\}$ be the $t$-MIPPC$(n,M,q)$ defined on $Q = \{0,1, \ldots, q-1\}$,
and ${\cal E} = \{{\bf e}_1, {\bf e}_2, \ldots, {\bf e}_{q}\}$, where ${\bf e}_i$ is the $i$-th column identity vector,
i.e., all its coordinates are $0$ except the $i$-th one being $1$.
Let $f: Q \longrightarrow {\cal E}$ be the bijective mapping such that $f(i) = {\bf e}_{i+1}$.
For any codeword ${\bf c} = ({\bf c}(1), {\bf c}(2), \ldots, {\bf c}(n))^{T} \in {\cal C}$,
we define $f({\bf c}) = (f({\bf c}(1)), f({\bf c}(2)), \ldots, f({\bf c}(n)))$.
Obviously, $f({\bf c})$ is a binary column vector of length $nq$.
We define a new $(nq, M, 2)$ code ${\cal F} = \{f({\bf c}_1), f({\bf c}_2), \ldots, f({\bf c}_M)\}$.
We are going to show that ${\cal F}$ is in fact a multimedia $t$-IPP code.

Consider any $S \subseteq {\cal F}(1) \times \cdots \times {\cal F}(nq)$
with ${\cal P}_{t}(S) = \{{\cal F}_1, \ldots, {\cal F}_r\} \neq \emptyset$.
Each ${\cal F}_i$ corresponds to a subcode ${\cal C}_i \subseteq {\cal C}$ such that $|{\cal C}_i| \le t$,
where ${\cal F}_i = \{f({\bf c}) \ | \ {\bf c} \in {\cal C}_i\}$.
Since ${\sf desc}({\cal F}_1) = {\sf desc}({\cal F}_2) = \cdots = {\sf desc}({\cal F}_r)$,
we immediately have ${\sf desc}({\cal C}_1) = {\sf desc}({\cal C}_2) = \cdots = {\sf desc}({\cal C}_r)$.
Since ${\cal C}$ is a $t$-MIPPC$(n,M,q)$, we have $\bigcap_{i = 1}^{r}{\cal C}_i \neq \emptyset$.
Let ${\bf c} \in \bigcap_{i = 1}^{r}{\cal C}_i$, then ${\bf c} \in {\cal C}_i$ for any $1 \leq i \leq r$,
which implies $f({\bf c}) \in {\cal F}_i$ for any $1 \leq i \leq r$, and thus $f({\bf c}) \in \bigcap_{i = 1}^{r}{\cal F}_i$.
Therefore, $\bigcap_{i = 1}^{r}{\cal F}_i \neq \emptyset$. This completes the proof.\end{IEEEproof}

The above theorem stimulates us to investigate $q$-ary multimedia $t$-IPP codes.
In the remaining parts of this paper, we will focus on the properties on the constructions of $q$-ary multimedia $t$-IPP codes.
\section{A General Upper Bound on the Code Size } %
\label{tESC}                                                                  %

Bipartite graphs are extensively used in modern coding theory, see, for example, \cite{ETV, Tan}.
In this section, we use bipartite graphs to derive an upper bound on the size of a $t$-MIPPC$(n,M,q)$.

Let $G(X,Y) = G(u,v)$ be a bipartite graph on $u$ vertices in the class $X$ and $v$ vertices in the class $Y$.
Without loss of generality, we may assume that $u \geq v$.
Let $e(G)$ denote the number of edges of $G$, that is, the size of $G$.
The girth of $G$ is the length of a shortest cycle in $G$.
It is well known that any bipartite graph is free of odd cycles.

\begin{lemma}
\label{bound3333}{\rm (\cite{Lam1,Lam2})}
If a bipartite graph $G(u,v)$ contains no cycle of length less than or equal to $2l$, where $u \geq v$, then
\[
e(G) \leq \left\{\begin{array}{ll}
(uv)^{\frac{l+1}{2l}}+ c(u+v), & l \mbox{ is odd}, \\[2pt]
v^{\frac{1}{2}}u^{\frac{l+2}{2l}} + c(u+v), & l \mbox{ is even}, \\[2pt]
\end{array}
\right.
\]
where $c$ is a constant depending only on $l$.
\end{lemma}

An application of Lemma \ref{bound3333} is the following theorem.

\begin{theorem}
\label{bound4444}
$M(t,n,q) \leq q^{\frac{n}{2}}(q^{\frac{n}{2t}} + 2c)$ if $n$ is even, and
\[M(t,n,q) \leq \left\{\begin{array}{ll}
q^{\frac{n}{2}}(q^{\frac{n+1}{2t}} + c(q^{\frac{1}{2}}+ q^{-\frac{1}{2}})), & t \mbox{ is even}, \\[2pt]
q^{\frac{n}{2}}(q^{\frac{n}{2t}} + c(q^{\frac{1}{2}}+ q^{-\frac{1}{2}})), & t \mbox{ is odd} \\[2pt]
\end{array}
\right.\]
if $n$ is odd, where $c$ is a constant depending only on $t$.
\end{theorem}
\begin{IEEEproof} Let $\cal C$ be a $t$-MIPPC$(n,M,q)$ defined on $Q$.
We prove this theorem in two cases.

If $n$ is even, we construct a bipartite graph $G(q^{\frac{n}{2}},q^{\frac{n}{2}})$ as follows. Let $X = Y = Q^{\frac{n}{2}}$.
An edge connects ${\bf a} \in X$ and ${\bf b} \in Y$ if and only if $({\bf a}, {\bf b})^{T} \in \cal C$. Obviously, $M = e(G)$.
Suppose that there exists a $2t_0$-cycle in $G$, where $ 2 \leq t_0 \leq t$.
Let $({\bf a}_1, {\bf b}_1, {\bf a}_2, {\bf b}_2, \ldots, {\bf a}_{t_0}, {\bf b}_{t_0})$
be the  $2t_0$-cycle, where ${\bf a}_i$, $1 \leq i \leq t_0$, are distinct vertices in $X$,
and ${\bf b}_i$, $1 \leq i \leq t_0$, are distinct vertices in $Y$.
Then $({\bf a}_i, {\bf b}_i)^{T} \in \cal C$ for $1 \leq i \leq t_0$, and
$({\bf a}_1, {\bf b}_{t_0})^{T}, ({\bf a}_i, {\bf b}_{i-1})^{T}\in \cal C$ for $2 \leq i \leq t_0$.
Let ${\cal C}_1 =\{({\bf a}_i, {\bf b}_i)^{T} \ | \ 1 \leq i \leq t_0\}$,
${\cal C}_2 = \{({\bf a}_1, {\bf b}_{t_0})^{T}\} \bigcup \{({\bf a}_i, {\bf b}_{i-1})^{T} \ | \ 2 \leq i \leq t_0\}$.
Then ${\sf desc}({\cal C}_1) = {\sf desc}({\cal C}_2)$, but ${\cal C}_1 \bigcap {\cal C}_2 = \emptyset$,
a contradiction to the fact that ${\cal C}$ is a $t$-MIPPC$(n,M,q)$.
So $G$ contains no cycle of length less than or equal to $2t$.
The conclusion then comes from Lemma \ref{bound3333}.

If $n$ is odd, we construct a bipartite graph $G(q^{\frac{n+1}{2}},q^{\frac{n-1}{2}})$ with $X = Q^{\frac{n+1}{2}}, Y = Q^{\frac{n-1}{2}}$.
Similarly, we can show that $G$ contains no cycle of length less than or equal to $2t$,
and the conclusion follows by Lemma \ref{bound3333}.
\end{IEEEproof}

\section{multimedia $3$-IPP Codes} %
\label{3ESC}                                            %

In order to derive a tight bound on the size of a multimedia $3$-IPP code,
we present a combinatorial characterization of multimedia $3$-IPP codes.

For any $(n,M,q)$ code ${\cal C}$ on $Q= \{0,1, \ldots, q-1\}$, Cheng {\it et al.} \cite{CJM} defined
the following column vector sets ${\cal  A}_{i}^{j}$ for $i \in Q$ and $1 \leq j \leq n$:
\begin{eqnarray*}
 {\cal  A}_{i}^{j} = \{({\bf c}(1), \ldots, {\bf c}(j-1), {\bf c}(j+1), \ldots, {\bf c}(n))^{T} \ | \\
({\bf c}(1), \ldots, {\bf c}(n))^{T} \in {\cal C}, {\bf c}(j) = i\}.
\end{eqnarray*}

We first prove the following lemma on $\overline{2}$-separable codes.

\begin{lemma}
\label{2barSC}
Let $\cal C$ be a $(2,M,q)$ code.
Then $\cal C$ is a $\overline{2}$-SC$(2,M,q)$ if and only if
$|{\cal A}_{a_1}^1 \bigcap {\cal A}_{a_2}^1| \leq 1$ holds in ${\cal C}$  for any distinct elements $a_1, a_2 \in Q$.
\end{lemma}
\begin{IEEEproof} The necessity is in fact a special case of Theorem 3.9 in \cite{CJM}.
Let $\cal C$ be a $\overline{2}$-SC$(2,M,q)$.
Assume that there exist distinct elements $a_1, a_2\in Q$ satisfying $|{\cal  A}_{a_1}^1 \bigcap {\cal A}_{a_2}^1| \geq 2$.
Suppose $b_1, b_2\in{\cal A}_{a_1}^1 \bigcap {\cal A}_{a_2}^1$, $b_1\neq b_2$.
Then $(a_1,b_1)^T$, $(a_1,b_2)^T$, $(a_2,b_1)^T$, $(a_2,b_2)^T \in \cal C$.
Let ${\cal C}_1 = \{(a_1,b_1)^T$, $(a_2,b_2)^T\}$ and ${\cal C}_2 = \{(a_1,b_2)^T$, $(a_2,b_1)^T\}$.
Then ${\cal C}_1 \neq {\cal C}_2$ and ${\sf desc}({\cal C}_1) = {\sf desc}({\cal C}_2)$,
a contradiction to the definition of a $\overline{2}$-SC$(2,M,q)$.

Now we consider its sufficiency. Suppose that $|{\cal A}_{a_1}^1 \bigcap {\cal A}_{a_2}^1|$ $ \leq 1$
holds in ${\cal C}$ for any distinct elements $a_1, a_2 \in Q$, but $\cal C$ is not a $\overline{2}$-SC$(2,M,q)$.
This implies that there exist ${\cal C}_1,{\cal C}_2 \subseteq {\cal C}$, ${\cal C}_1 \neq {\cal C}_2$,
$|{\cal C}_1| \leq 2$ and $|{\cal C}_2| \leq 2$, such that ${\sf desc}({\cal C}_1) = {\sf desc}({\cal C}_2)$.

Let ${\cal C}_1 = \{{\bf c}_1,{\bf c}_2\}$, ${\cal C}_2 = \{{\bf c}_3,{\bf c}_4\}$, ${\cal C}_1 \neq {\cal C}_2$,
and ${\bf c}_i = (a_i,b_i)^{T}$ for $1 \leq i \leq 4$.
We remark here that we allow ${\bf c}_1 = {\bf c}_2$ or ${\bf c}_3 = {\bf c}_4$.
Since ${\sf desc}({\cal C}_1) = {\sf desc}({\cal C}_2)$, then ${\cal C}_1(1)={\cal C}_2(1)$ and ${\cal C}_1(2)={\cal C}_2(2)$.
This implies that $a_1 = a_2$ (or $a_3 = a_4$) if and only if $a_1 = a_2 = a_3 = a_4$,
and $b_1 = b_2$ (or $b_3 = b_4$) if and only if $b_1 = b_2 = b_3 = b_4$.

Now, if $a_1 = a_2$, then $a_1 = a_2 = a_3 = a_4$. Since ${\cal C}_1 \neq {\cal C}_2$, we have $b_1 \neq b_2$.
By the fact that ${\cal C}_1(2) = {\cal C}_2(2)$, we have $\{b_1,b_2\} = \{b_3,b_4\}$,
and therefore ${\cal C}_1 = {\cal C}_2$, a contradiction.
On the other hand, if $a_1 \neq a_2$, then $a_3 \neq a_4$.
Clearly, $b_1 \neq b_2$, otherwise we can use a similar argument to conclude that ${\cal C}_1 = {\cal C}_2$.
Now, we have $\{a_1,a_2\} = \{a_3,a_4\}$ and $\{b_1,b_2\} = \{b_3,b_4\}$ as set equalities.
Without loss of generality, we may assume $a_1=a_3$ and $a_2=a_4$.
In this case, if $b_1=b_3$, then $b_2=b_4$, and thus ${\cal C}_1 = {\cal C}_2$, a contradiction.
Therefore, $b_1=b_4$ and $b_2 = b_3$, which implies that
${\cal A}_{a_1}^1 \bigcap {\cal A}_{a_2}^1 = \{b_1,b_2\}$, again a contradiction.
This completes the proof.
\end{IEEEproof}

Now we turn our attention to multimedia $3$-IPP codes.

\begin{lemma}
\label{neccond}
Let $\cal C$ be a $3$-MIPPC$(n,M,q)$ code defined on $Q = \{0,1, \ldots, q-1\}$. Then
\begin{itemize}
\item[(I)] $|{\cal A}_{a_1}^1 \bigcap {\cal A}_{a_2}^1| \leq 1$ always holds for any distinct elements $a_1, a_2 \in Q$;
\item[(II)] There do not exist distinct elements $a_1, a_2, a_3 \in Q$ and distinct vectors ${\bf b}_1, {\bf b}_2, {\bf b}_3 \in Q^{n-1}$
such that ${\bf b}_1, {\bf b}_2 \in {\cal A}_{a_1}^1$, ${\bf b}_2, {\bf b}_3 \in {\cal A}_{a_2}^1$, ${\bf b}_1, {\bf b}_3 \in {\cal A}_{a_3}^1$.
\end{itemize}
\end{lemma}
\begin{IEEEproof}
\begin{itemize}
\item[(I)] If there exist distinct elements $a_1,a_2 \in Q$ satisfying that $|{\cal  A}_{a_1}^1 \bigcap {\cal A}_{a_2}^1| \geq 2$,
say ${\bf b}_1 \neq {\bf b}_2 \in {\cal A}_{a_1}^1 \bigcap {\cal A}_{a_2}^1$,
then $(a_1, {\bf b}_1)^T, (a_1, {\bf b}_2)^T, (a_2, {\bf b}_1)^T, (a_2, {\bf b}_2)^T  \in {\cal C}$.
Let ${\cal C}_1 = \{(a_1, {\bf b}_1)^T, (a_2, {\bf b}_2)^T\}$ and ${\cal C}_2 = \{(a_1, {\bf b}_2)^T, (a_2, {\bf b}_1)^T\}$.
Then ${\sf desc}({\cal C}_1) = {\sf desc}({\cal C}_2)$, but ${\cal C}_{1} \bigcap {\cal C}_{2} = \emptyset$,
 a contradiction to the definition of a $3$-MIPPC$(n,M,q)$.
\item[(II)] If there exist distinct elements $a_1, a_2, a_3 \in Q$ and distinct vectors ${\bf b}_1, {\bf b}_2, {\bf b}_3 \in Q^{n-1}$
such that ${\bf b}_1, {\bf b}_2 \in {\cal A}_{a_1}^1$, ${\bf b}_2, {\bf b}_3 \in {\cal A}_{a_2}^1$,
${\bf b}_1, {\bf b}_3 \in {\cal A}_{a_3}^1$,  then
$(a_1,{\bf b}_1)^T$, $(a_1,{\bf b}_2)^T$, $(a_2,{\bf b}_2)^T$, $(a_2,{\bf b}_3)^T$, $(a_3,{\bf b}_1)^T$, $(a_3,{\bf b}_3)^T \in \cal C$.
Let ${\cal C}_1$ $=\{(a_1,{\bf b}_1)^T$, $ (a_2,{\bf b}_2)^T$, $(a_3,{\bf b}_3)^T\}$,
${\cal C}_2 =\{(a_1, {\bf b}_2)^T$, $(a_2,{\bf b}_3)^T, (a_3, {\bf b}_1)^T\}$.
Then ${\sf desc}({\cal C}_1) = {\sf desc}({\cal C}_2)$, but ${\cal C}_{1} \bigcap {\cal C}_{2} = \emptyset$,
a contradiction to the definition of a $3$-MIPPC$(n,M,q)$.
\end{itemize}
\end{IEEEproof}

It is of interest to see that the converse of Lemma \ref{neccond} is true when $n = 2$.

\begin{lemma}
\label{sufcond}
Let $\cal C$ be a $(2,M,q)$ code defined on $Q=\{0,1, \ldots, q-1\}$.
If $\cal C$ satisfies the following two conditions:
\begin{itemize}
\item[(I)] $|{\cal A}_{a_1}^1 \bigcap {\cal A}_{a_2}^1| \leq 1$ always holds for any distinct elements $a_1,a_2 \in Q$;
\item[(II)] There do not exist distinct elements $a_1, a_2, a_3 \in Q$ and distinct elements $b_1,  b_2, b_3 \in Q$,
such that $b_1, b_2 \in {\cal A}_{a_1}^1$, $b_2, b_3 \in {\cal A}_{a_2}^1$, $b_1, b_3 \in {\cal A}_{a_3}^1$.
\end{itemize}
Then $\cal C$ is a $3$-MIPPC$(2,M,q)$.
\end{lemma}
\begin{IEEEproof} Suppose $\cal C$ satisfies conditions (I) and (II). We prove this lemma in three steps.

(1) At first, we prove that if there exist ${\cal C}_1,{\cal C}_2 \subseteq {\cal C}$, ${\cal C}_1 \neq {\cal C}_2$,
$|{\cal C}_1| \leq 3$, $|{\cal C}_2| \leq 3$, satisfying ${\sf desc}({\cal C}_1) = {\sf desc}({\cal C}_2)$, then
${\cal C}_1$ and ${\cal C}_2$ should be of one of the following three types:
\begin{eqnarray*}
\hbox{Type {\bf I}:}\ \ \ \
\begin{array}{c}
{\bf c}_1 \ \ \ {\bf c}_2 \ \ \  {\bf c}_3 \\
\left( \begin{array}{ccc}
 a_1 & a_2 &  a_1 \\
b_1 & b_2 &  b_2
\end{array} \right)
\end{array} ,
\end{eqnarray*}
where ${\cal C}_1 = \{{\bf c}_1, {\bf c}_2\}$, ${\cal C}_2 = \{{\bf c}_1, {\bf c}_2, {\bf c}_3\}$, $a_1 \neq a_2$, $b_1 \neq b_2$;
\begin{eqnarray*}
\hbox{Type {\bf II}:}\ \ \ \
\begin{array}{c}
  {\bf c}_1  \ \ \  {\bf c}_2  \ \ \  {\bf c}_3  \ \ \  {\bf c}_4 \\
\left(\begin{array}{cccc}
  a_1 & a_2 &  a_3 & a_1  \\
  b_1 & b_1 &  b_3 & b_3 \\
\end{array} \right)
\end{array} ,
\end{eqnarray*}
where ${\cal C}_1 = \{{\bf c}_1,{\bf c}_2,{\bf c}_3 \}$, ${\cal C}_2= \{{\bf c}_2, {\bf c}_3, {\bf c}_4 \}$,
$a_{k_1} \neq a_{k_2}$, $1 \leq k_1 < k_2 \leq 3$, $b_1 \neq b_3$;
\begin{eqnarray*}
\hbox{Type {\bf III}:}\ \ \ \
\begin{array}{c}
  {\bf c}_1 \ \ \  {\bf c}_2  \ \ \  {\bf c}_3  \ \ \  {\bf c}_4 \\
  \left(\begin{array}{cccc}
  a_1 & a_1 &  a_3 & a_3 \\
  b_1 & b_2 &  b_3 & b_1 \\
\end{array} \right)
\end{array} ,
\end{eqnarray*}
where ${\cal C}_1 = \{{\bf c}_1, {\bf c}_2, {\bf c}_3 \}$, ${\cal C}_2= \{ {\bf c}_2, {\bf c}_3, {\bf c}_4 \}$,
$a_1 \neq a_3$, $b_{k_1} \neq b_{k_2}$, $1 \leq k_1 < k_2 \leq 3$.

(1.1) If $|{\cal C}_1| \leq 2$, $|{\cal C}_2| \leq 2$, then ${\cal C}$ is not a $\overline{2}$-SC$(2, M, q)$.
However, according to condition (I) and Lemma \ref{2barSC}, $\cal C$ is a $\overline{2}$-SC$(2, M, q)$, a contradiction.
So this case is impossible.

(1.2) If $|{\cal C}_1| = 1$, $|{\cal C}_2| = 3$, let ${\cal C}_1 = \{{\bf c}_1\}$, ${\cal C}_2 = \{{\bf c}_2, {\bf c}_3, {\bf c}_4\}$,
where ${\bf c}_i = (a_i, b_i)^T$, $1 \leq i \leq 4$.
Then $a_1 = a_2 = a_3 =a_4$ and $b_1 = b_2 = b_3 = b_4$ according to ${\sf desc}({\cal C}_1) = {\sf desc}({\cal C}_2)$,
which implies ${\bf c}_1 = {\bf c}_2 = {\bf c}_3 = {\bf c}_4$, a contradiction. So this case is not possible either.

(1.3) Consider the case $|{\cal C}_1| = 2$, $|{\cal C}_2| =3$.
Let $|{\cal C}_1| = \{{\bf c}_1, {\bf c}_2\}$, $|{\cal C}_2| = \{{\bf c}_3, {\bf c}_4, {\bf c}_5\}$,
where ${\bf c}_i = (a_i, b_i)^T$, $1 \leq i \leq 5$.

(1.3.A) If $a_1 = a_2$, then $a_3 = a_4 = a_5 = a_1$. Since $\{b_1, b_2\} = \{b_3, b_4, b_5\}$,
there must be two identical elements in $\{b_3, b_4, b_5\}$. We may assume $b_3 = b_4$.
Then ${\bf c}_3 = {\bf c}_4$, a contradiction. So this case is impossible.

(1.3.B) If $a_1 \neq a_2$, since ${\sf desc}({\cal C}_1) = {\sf desc}({\cal C}_2)$,
then $a_3, a_4, a_5 \in \{a_1, a_2\}$ and $b_3, b_4, b_5 \in \{b_1, b_2\}$.
Without loss of generality, we may assume that $a_3 = a_4 = a_1$ and  $a_5 = a_2$.
Then $b_3 \neq b_4$, otherwise, ${\bf c}_3 = {\bf c}_4$, a contradiction.
Since $b_3, b_4 \in \{b_1, b_2\}$, then $b_1 \neq b_2$ and we may assume that $b_3 = b_1$ and $b_4 = b_2$.
\begin{eqnarray*}
\begin{array}{c}
  {\bf c}_1 \ \ \ {\bf c}_2 \ \big| \ {\bf c}_3  \ \ \ {\bf c}_4 \ \  {\bf c}_5 \\
\left(\begin{array}{cc|ccc}
  a_1 & a_2 & a_1  & a_1 & a_2\\
  b_1 & b_2 & b_1  & b_2 &  \\
  \end{array}\right)
  \end{array}
\end{eqnarray*}

If $b_5 = b_1$, then $b_1, b_2 \in {\cal A}_{a_1}^1 \bigcap {\cal A}_{a_2}^1$,
that is, $|{\cal A}_{a_1}^1 \bigcap {\cal A}_{a_2}^1| \geq 2$, a contradiction to condition (I).
So this case is impossible.

If $b_5 = b_2$, then
\begin{eqnarray*}
\begin{array}{c}
  {\bf c}_1 \ \ \  {\bf c}_2\ \big| \  {\bf c}_3  \ \ \ {\bf c}_4 \ \ \ {\bf c}_5 \\
\left(\begin{array}{cc|ccc}
  a_1 & a_2 & a_1  & a_1 & a_2\\
  b_1 & b_2 & b_1  & b_2 & b_2\\
  \end{array}\right)
  \end{array},
\end{eqnarray*}
that is,
\begin{eqnarray*}
\begin{array}{c}
  {\bf c}_1 ({\bf c}_3) \  {\bf c}_2 ({\bf c}_5) \ \ {\bf c}_4  \\
\left(\begin{array}{ccc}
\  \  a_1\   &\ \ a_2\  & \  a_1 \\
\  \ b_1 \   &\ \  b_2 \  &\   b_2 \\
  \end{array}\right)
  \end{array}.
\end{eqnarray*}
So ${\cal C}_1$ and ${\cal C}_2$ are of type {\bf I}.

(1.4) Consider the case $|{\cal C}_1| = 3$, $|{\cal C}_2| =3$.
Let ${\cal C}_1 = \{{\bf c}_1, {\bf c}_2, {\bf c}_3\}$, ${\cal C}_2 = \{{\bf c}_4, {\bf c}_5, {\bf c}_6\}$,
where ${\bf c}_i = (a_i, b_i)^T$, $1 \leq i \leq 6$.

(1.4.A) If $a_1=a_2=a_3$ or $b_1=b_2=b_3$, then ${\cal C}_1={\cal C}_2$, a contradiction. So this case is impossible.

(1.4.B) Consider the case $a_1 = a_2$ and  $a_3 \neq a_1$.
Then $b_1 \neq b_2$, otherwise, ${\bf c}_1 = {\bf c}_2$, a contradiction.

(1.4.B.a) Suppose $b_1=b_3$. Since $a_3 \in \{a_4, a_5, a_6\}$, we may assume $a_4 = a_3$.
Then $b_4= b_1$, otherwise, $b_4= b_2$, which implies $b_1, b_2 \in {\cal A}_{a_1}^1 \bigcap {\cal A}_{a_3}^1$,
a contradiction to condition (I).
\begin{eqnarray*}
\begin{array}{c}
{\bf c}_1 \ \ \ {\bf c}_2 \ \ ~ {\bf c}_3  \begin{footnotesize}~~\end{footnotesize} \big| \ {\bf c}_4 \ \ \ {\bf c}_5 \ \ \ {\bf c}_6 \ \\
\left(\begin{array}{ccc|ccc}
  a_1 & a_1 & a_3  & a_3 & \ ~ ~  & \ ~~  \\
  b_1 & b_2 & b_1  & b_1 & \ ~ ~  &\ ~~ \\
  \end{array}\right)
  \end{array}
\end{eqnarray*}
Now we consider ${\bf c}_5$ and ${\bf c}_6$.
If $a_5 = a_3$ or $a_6 = a_3$, similarly, we can show that $b_5= b_1$ or $b_6= b_1$, respectively,
which implies ${\bf c}_5 = {\bf c}_4$ or ${\bf c}_6 = {\bf c}_4$, respectively, a contradiction. So $a_5 = a_6 = a_1$.
Then $b_5 \neq b_6$, otherwise, ${\bf c}_5 = {\bf c}_6$, a contradiction.
Since $b_5, b_6 \in \{b_1, b_2\}$,  we may assume that $b_5 = b_1$, $b_6 = b_2$.
\begin{eqnarray*}
\begin{array}{c}
{\bf c}_1 \ \ \  {\bf c}_2 \ \ \  {\bf c}_3  \ \big| \ {\bf c}_4 \ \ \  {\bf c}_5 \ \ \ {\bf c}_6  \\
\left(\begin{array}{ccc|ccc}
  a_1 & a_1 & a_3  & a_3 & a_1 & a_1\\
  b_1 & b_2 & b_1  & b_1 & b_1 & b_2 \\
  \end{array}\right)
  \end{array}
\end{eqnarray*}
Then ${\cal C}_1 = {\cal C}_2$, a contradiction. So this case is impossible.

(1.4.B.b) Suppose $b_i \neq b_j$, $1 \leq i < j \leq 3$.
Since $\{b_1, b_2,$ $ b_3\} = \{b_4, b_5, b_6\}$, we may assume that $b_4 = b_1, b_5 = b_2, b_6 = b_3$.
\begin{eqnarray*}
\begin{array}{c}
\ {\bf c}_1 \ \ \ {\bf c}_2 \ \  \begin{footnotesize}~\end{footnotesize} {\bf c}_3  \begin{footnotesize}~\end{footnotesize} \big|\begin{footnotesize}~\end{footnotesize} {\bf c}_4 \ \ \ {\bf c}_5 \ \ \ {\bf c}_6  \\
\left(\begin{array}{ccc|ccc}
  a_1 & a_1 & a_3  &  &  & \\
  b_1 & b_2 & b_3  & b_1 & b_2 & b_3 \\
  \end{array}\right)
  \end{array}
\end{eqnarray*}

It is impossible that $(a_4, a_5) = (a_1, a_1)$. Otherwise, $a_6 = a_3$, which implies ${\cal C}_1 = {\cal C}_2$, a contradiction.

It is not possible either that $(a_4, a_5) = (a_3, a_3)$. Otherwise, $b_1, b_2 \in {\cal A}_{a_1}^1 \bigcap {\cal A}_{a_3}^1$,
a contradiction to condition (I).

If $(a_4, a_5) = (a_1, a_3)$, then
\begin{eqnarray*}
\begin{array}{c}
  {\bf c}_1 \ \ \  {\bf c}_2 \ \ \  {\bf c}_3  \ \big|\begin{footnotesize}~\end{footnotesize}  {\bf c}_4 \ \ \ {\bf c}_5 \ \ \ {\bf c}_6  \\
\left(\begin{array}{ccc|ccc}  a_1 & a_1 & a_3  & a_1 & a_3 & \\
  b_1 & b_2 & b_3  & b_1 & b_2 & b_3 \\
  \end{array}\right)
  \end{array}.
\end{eqnarray*}
We should have $a_6=a_3$. Otherwise, $a_6 = a_1$, then $b_2, b_3 \in {\cal A}_{a_1}^1 \bigcap {\cal A}_{a_3}^1$,
a contradiction to condition (I). So
\begin{eqnarray*}
\begin{array}{c}
  {\bf c}_2  \ \ {\bf c}_1 ({\bf c}_4) \ \  {\bf c}_3 ({\bf c}_6) \ \ {\bf c}_5 \\
\left(\begin{array}{cccc}
\ a_1\  & \ a_1 \ &\  a_3\  &\ a_3\  \\
\ b_2\  &\ b_1 \ &\  b_3\  &\  b_2\   \\
  \end{array}\right)
  \end{array},
\end{eqnarray*}
and therefore, ${\cal C}_1$ and ${\cal C}_2$ are of type {\bf III}.

Similarly, if $(a_4, a_5) = (a_3, a_1)$, we can show that ${\cal C}_1$ and ${\cal C}_2$ are of type {\bf III}.

(1.4.C) Consider the case $a_i \neq a_j$, $1 \leq i < j \leq 3$.
Since $\{a_1, a_2, a_3\} = \{a_4, a_5, a_6\}$, we may assume that $a_4 = a_1, a_5 = a_2, a_6 = a_3$.

(1.4.C.a) Suppose $b_1 = b_2$ and  $b_3 \neq b_1$.
\begin{eqnarray*}
\begin{array}{c}
{\bf c}_1 \ \ \ {\bf c}_2 \ \ \ {\bf c}_3 \ \big| \ {\bf c}_4 \ \ \ {\bf c}_5 \ \ \ {\bf c}_6\\
\left(\begin{array}{ccc|ccc}
 a_1 & a_2 & a_3  & a_1 & a_2 & a_3\\
  b_1 & b_1 & b_3  &  &  &  \\
  \end{array}\right)
  \end{array}
\end{eqnarray*}

It is impossible that $(b_4, b_5) = (b_1, b_1)$. Otherwise, $b_6 = b_3$, which implies ${\cal C}_1 = {\cal C}_2$, a contradiction.

It is not possible either that $(b_4, b_5) = (b_3, b_3)$.
Otherwise, $b_1, b_3 \in {\cal A}_{a_1}^1 \bigcap {\cal A}_{a_2}^1$, a contradiction to condition (I).

Suppose $(b_4, b_5) = (b_1, b_3)$.
\begin{eqnarray*}
\begin{array}{c}
  {\bf c}_1 \ \ \ {\bf c}_2 \ \ \ {\bf c}_3 \ \big| \ {\bf c}_4 \ \ \ {\bf c}_5 \ \ \ {\bf c}_6  \\
\left(\begin{array}{ccc|ccc}
  a_1 & a_2 & a_3  & a_1 & a_2 & a_3\\
  b_1 & b_1 & b_3  & b_1 & b_3 &  \\
  \end{array}\right)
  \end{array}
\end{eqnarray*}
Then $b_6 = b_3$. Otherwise, $b_6 = b_1$, then $b_1, b_3 \in {\cal A}_{a_2}^1 \bigcap {\cal A}_{a_3}^1$,
a contradiction to condition (I). So
\begin{eqnarray*}
\begin{array}{c}
  {\bf c}_2  \ \  {\bf c}_1 ({\bf c}_4) \ \  {\bf c}_3 ({\bf c}_6) \ \ {\bf c}_5 \\
\left(\begin{array}{cccc}
\ a_2\  &\ a_1\ &\  a_3\ &\ a_2\  \\
\  b_1\ &\ b_1\ &\  b_3\ &\ b_3\ \\
  \end{array}\right)
  \end{array}
\end{eqnarray*}
and thus ${\cal C}_1$ and ${\cal C}_2$ are of type {\bf II}.

SimilarIy, if $(b_4, b_5) = (b_3, b_1)$, we can derive that ${\cal C}_1$ and ${\cal C}_2$ are of type {\bf II}.

(1.4.C.b) Suppose $b_i \neq b_j$, $1 \leq i < j \leq 3$.
\begin{eqnarray*}
\begin{array}{c}
{\bf c}_1 \ \ \ {\bf c}_2 \ \ \ {\bf c}_3  \ \big| \  {\bf c}_4\ \ \ {\bf c}_5\ \ \  {\bf c}_6  \\ \left(\begin{array}{ccc|ccc}
a_1 & a_2 & a_3  & a_1 & a_2 & a_3\\
  b_1 & b_2 & b_3  &  &  &  \\
  \end{array}\right)
  \end{array}
\end{eqnarray*}

It is impossible that $(b_4, b_5,b_6) = (b_1, b_2,b_3)$. Otherwise, ${\cal C}_1 = {\cal C}_2$, a contradiction.

It is impossible that $(b_4, b_5,b_6) = (b_1, b_3,b_2)$.
Otherwise, $b_2, b_3 \in {\cal A}_{a_2}^1 \bigcap {\cal A}_{a_3}^1$, a contradiction to condition (I).

It is impossible that $(b_4, b_5,b_6) = (b_2, b_1,b_3)$.
Otherwise, $b_1, b_2 \in {\cal A}_{a_1}^1 \bigcap {\cal A}_{a_2}^1$, a contradiction to condition (I).

It is impossible that $(b_4, b_5,b_6) = (b_2, b_3,b_1)$.
Otherwise, $b_1, b_2 \in {\cal A}_{a_1}^1$, $b_2, b_3 \in  {\cal A}_{a_2}^1$, $b_1, b_3 \in {\cal A}_{a_3}^1$,
a contradiction to condition (II).

It is impossible that $(b_4, b_5,b_6) = (b_3, b_1,b_2)$.
Otherwise, $b_1, b_3 \in {\cal A}_{a_1}^1$, $b_1, b_2 \in  {\cal A}_{a_2}^1$, $b_2, b_3 \in {\cal A}_{a_3}^1$,
a contradiction to condition (II).

Finally, it is not possible either that $(b_4, b_5,b_6) = (b_3, b_2,b_1)$.
Otherwise, $b_1, b_3 \in {\cal A}_{a_1}^1 \bigcap {\cal A}_{a_3}^1$, a contradiction to condition (I).

(2) Now we prove that $|{\cal P}_{3}(S)| \leq 2$ for any $S \subseteq {\cal C}(1) \times {\cal C}(2)$.
Assume that there exists $S \subseteq {\cal C}(1) \times {\cal C}(2)$ such that $|{\cal P}_{3}(S)| \geq 3$.
Let ${\cal C}_1,{\cal C}_2, {\cal C}_3 \in {\cal P}_{3}(S)$ be three distinct sub-codes of $\cal C$.
According to (1), ${\sf desc}({\cal C}_i) = {\sf desc}({\cal C}_j)$ implies ${\cal C}_i$ and ${\cal C}_j$
are of one of the three types described in (1), where $1 \leq i < j\leq 3$.

(2.1) If there exists an index $i$, $1 \leq i \leq 3$, such that $|{\cal C}_i| = 2$,
without loss of generality, we may assume $|{\cal C}_1| = 2$.
Then ${\cal C}_1$ and ${\cal C}_2$ are of type {\bf I}, ${\cal C}_1$ and ${\cal C}_3$ are of type {\bf I}.
We may assume that ${\cal C}_1 = \{{\bf c}_1, {\bf c}_2\}$, ${\cal C}_2= \{{\bf c}_1, {\bf c}_2, {\bf c}_3\}$,
and ${\cal C}_3= \{{\bf c}_1, {\bf c}_2, {\bf c}_4\}$, where ${\bf c}_i = (a_i, b_i)^T$, $1 \leq i \leq 4$.
According to type I, ${\bf c}_3, {\bf c}_4 \in \{(a_1, b_2)^T, (a_2, b_1)^T\}$.
Clearly ${\bf c}_3 \neq {\bf c}_4$, otherwise ${\cal C}_2 = {\cal C}_3$, a contradiction.
Therefore, $b_1, b_2 \in {\cal A}_{a_1}^1 \bigcap {\cal A}_{a_2}^1$, which implies
$|{\cal A}_{a_1}^1 \bigcap {\cal A}_{a_2}^1| \geq 2$, a contradiction to condition (I).
So this case is impossible.

(2.2) Consider the case $|{\cal C}_i| = 3$ for all $1 \leq i \leq 3$.

(2.2.A) Suppose ${\cal C}_1$ and ${\cal C}_2$ are of type {\bf II}, ${\cal C}_1$ and ${\cal C}_3$ are of type {\bf II}.
Let ${\cal C}_1 = \{{\bf c}_1, {\bf c}_2, {\bf c}_3\}$, ${\cal C}_2= \{ {\bf c}_2, {\bf c}_3, {\bf c}_4\}$,
and ${\cal C}_3= \{{\bf c}_5, {\bf c}_6, {\bf c}_7\}$, where ${\bf c}_i = (a_i, b_i)^T$, $1 \leq i \leq 7$.
According to type {\bf II}, $a_{k_1} \neq a_{k_2}$, $1 \leq k_1 < k_2 \leq 3$, $b_1 \neq b_3$.
\begin{eqnarray*}
\begin{array}{c}
{\bf c}_1\ \ \ {\bf c}_2\ \ \ {\bf c}_3\ \ \ {\bf c}_4\ \ \ {\bf c}_5\ \ \ {\bf c}_6\ \ \ {\bf c}_7 \\
\left(\begin{array}{ccccccc}
a_1 & a_2 &  a_3 & a_1 &\ \ ~&\ \ ~ &\ \ ~ \\
b_1 & b_1 &  b_3 & b_3 &\ \ ~ &\ \ ~  &\ \ ~\\
\end{array}\right)
\end{array}
\end{eqnarray*}
Since ${\cal C}_1$ and ${\cal C}_3$ are of type {\bf II}, we have $|{\cal C}_1 \bigcap {\cal C}_3| = 2$.
Furthermore, because we require $b_1 \ne b_3$, we know ${\cal C}_1 \bigcap {\cal C}_3 \neq \{{\bf c}_1, {\bf c}_2\}$.

If ${\cal C}_1 \bigcap {\cal C}_3 = \{{\bf c}_1, {\bf c}_3\}$, we may assume ${\bf c}_5={\bf c}_1, {\bf c}_6 = {\bf c}_3$.
Then we should have ${\bf c}_7 = (a_2, b_3)^T$, and
\begin{eqnarray*}
\begin{array}{c}
{\bf c}_2 \ \ \ {\bf c}_1 ({\bf c}_5)\ \ {\bf c}_3  ({\bf c}_6)\ \ \ {\bf c}_7\ \ \  \ {\bf c}_4 \\
\left(\begin{array}{ccccc}
\ a_2\ &\ a_1\  &\ \ \ \ a_3\ &\ a_2\ &a_1\ \\
\ b_1\ &\ b_1\  &\ \ \ \ b_3\ & \ b_3\ & b_3\ \\
\end{array}\right)
\end{array},
\end{eqnarray*}
which implies $b_1, b_3 \in {\cal A}_{a_1}^1 \bigcap {\cal A}_{a_2}^1$, i.e.,
$|{\cal A}_{a_1}^1 \bigcap {\cal A}_{a_2}^1| \geq 2$, a contradiction to condition (I).
So this case is impossible.

If ${\cal C}_1 \bigcap {\cal C}_3 = \{{\bf c}_2, {\bf c}_3\}$, we may assume ${\bf c}_5={\bf c}_2, {\bf c}_6={\bf c}_3$.
Then ${\bf c}_7 = (a_1, b_3)^T = {\bf c}_4$, which implies ${\cal C}_2 = {\cal C}_3 $, a contradiction.
So this case is not possible either.

(2.2.B) Suppose ${\cal C}_1$ and ${\cal C}_2$ are of type {\bf III}, ${\cal C}_1$ and ${\cal C}_3$ are of type {\bf III}.
Similar to (2.2.A), we can prove this case is impossible.

(2.2.C) Suppose ${\cal C}_1$ and ${\cal C}_2$ are of type {\bf II}, ${\cal C}_1$ and ${\cal C}_3$ are of type {\bf III}.
Let ${\cal C}_1 = \{{\bf c}_1, {\bf c}_2, {\bf c}_3\}$, ${\cal C}_2= \{ {\bf c}_2, {\bf c}_3, {\bf c}_4\}$.
\begin{eqnarray*}
\begin{array}{c}
{\bf c}_1  \ \ \ {\bf c}_2\ \ \ {\bf c}_3\ \ \ {\bf c}_4 \\
\left(\begin{array}{cccc}
a_1 & a_2 &  a_3 & a_1 \\
  b_1 & b_1 &  b_3 & b_3 \\
\end{array}\right)
\end{array}
\end{eqnarray*}
Since $a_{k_1} \neq a_{k_2}$, $1 \leq k_1 < k_2 \leq 3$, it is impossible that ${\cal C}_1$ and ${\cal C}_3$ are of type {\bf III}.
So this case is not possible either.

Therefore, as we claimed earlier, $|{\cal P}_{3}(S)| \leq 2$ for any $S \subseteq {\cal C}(1) \times {\cal C}(2)$.

(3) Finally, the conclusion comes from (1), (2), and the fact that ${\cal C}_1 \bigcap {\cal C}_2 \neq \emptyset$
whenever ${\cal C}_1$ and ${\cal C}_2$ are of type {\bf I}, {\bf II}, or {\bf III}.
\end{IEEEproof}

Combining Lemma \ref{neccond} with Lemma \ref{sufcond}, we derive the main result of this section.

\begin{theorem}
\label{cond111}
Let $\cal C$ be a $(2,M,q)$ code defined on $Q=\{0,1, \ldots, q-1\}$.
Then $\cal C$ is a $3$-MIPPC$(2,M,q)$ if and only if it satisfies the following two conditions:
\begin{itemize}
\item[(I)] $|{\cal A}_{a_1}^1 \bigcap {\cal A}_{a_2}^1| \leq 1$ always holds for any distinct elements $a_1, a_2 \in Q$;
\item[(II)] There do not exist distinct elements $a_1, a_2, a_3 \in Q$ and distinct elements $b_1,  b_2, b_3 \in Q$
such that $b_1, b_2 \in {\cal A}_{a_1}^1$, $b_2, b_3 \in {\cal A}_{a_2}^1$, $b_1,b_3 \in {\cal A}_{a_3}^1$.
\end{itemize}
\end{theorem}

\section{Optimal $3$-MIPPC$(2,M,q)$s}            %

In Section \ref{tESC}, we have derived a general upper bound on the size of a $t$-MIPPC$(n,M,q)$.
Now, we are going to consider its optimality.

\begin{lemma}
\label{equiv222}
There exists a $3$-MIPPC$(2,M,q)$ if and only if
there exists a bipartite graph $G(q,q)$ of girth at least $8$ with $e(G)=M$.
\end{lemma}
\begin{IEEEproof} Suppose that there exists a $3$-MIPPC$(2,M,q)$, $\cal C$, defined on $Q$.
We construct a bipartite graph $G(q,q)$ as follows. Let $X = Q \times \{1\}$ and $Y = Q \times \{2\}$.
An edge is incident to $(a,1) \in X$ and $(b,2) \in Y$ if and only if $(a, b)^{T} \in \cal C$.
Then $e(G)=M$. We are going to show that $G$ has girth at least $8$.

Assume $G(q,q)$ contains a $4$-cycle, say $((a_1,1)$, $(b_1,2)$, $(a_2,1)$, $(b_2,2))$,
where $(a_i,1)$, $1 \leq i \leq 2$, are distinct elements of $X$, and $(b_i, 2)$, $1 \leq i \leq 2$, are distinct elements of $Y$.
Then $(a_1, b_1)^{T}, (a_2, b_1)^{T},$ $(a_2, b_2)^{T}, (a_1, b_2)^{T} \in \cal C$,
and thus $b_1, b_2 \in {\cal A}_{a_1}^1 \bigcap {\cal A}_{a_2}^1$, a contradiction to Theorem \ref{cond111}.
So this case is impossible.

Assume $G(q,q)$ contains a $6$-cycle, say $((a_1,1)$, $(b_1,2)$, $(a_2,1)$, $(b_2,2)$, $(a_3,1)$, $(b_3,2))$,
where $(a_i,1)$, $1\leq i\leq 3$, are  distinct elements of $X$, and $(b_i,2)$, $1 \leq i \leq 3$, are distinct elements of $Y$.
Then $(a_1,b_1)^{T}$, $(a_2,b_1)^{T}$, $(a_2, b_2)^{T}$, $(a_3,b_2)^{T}$, $(a_3,b_3)^{T}$, $(a_1, b_3)^{T} \in \cal C$,
and thus $b_1,b_3 \in {\cal A}_{a_1}^1$, $b_1, b_1 \in {\cal A}_{a_2}^1$, $b_2,b_3 \in {\cal A}_{a_3}^1$,
a contradiction to Theorem \ref{cond111}. So this case is not possible either.

Therefore, the bipartite graph $G(q,q)$ constructed above has girth at least $8$, with $e(G) = M$.

Conversely, for any bipartite graph $G(q,q)= G(X,Y)$ with girth at least $8$, we construct a $(2,M,q)$ code $\cal C$.
Let $Q = X$ and $f: Y \longrightarrow X$ be a bijective mapping.
A vector $(x, f(y))^{T} \in \cal C$ if and only if $\{x, y\}$ is an edge of G, where $x \in X$ and $y \in Y$.
Obviously, $\cal C$ is a $(2,M,q)$ code defined on $Q$ and $M=e(G)$.
Suppose that $\cal C$ is not a $3$-MIPPC$(2,M,q)$.
Then by Theorem \ref{cond111}, at least one of the following cases should happen.

(1) There exist distinct elements $x_1,x_2 \in Q$ such that $|{\cal A}_{x_1}^1$ $ \bigcap {\cal A}_{x_2}^1| \geq 2$.
In this case, we may assume $f(y_1) \neq f(y_2) \in {\cal A}_{x_1}^1 \bigcap {\cal A}_{x_2}^1$.
Then $y_1 \neq y_2$, and $(x_1, f(y_1))^{T}$, $(x_1, f(y_2))^{T}$, $(x_2, f(y_1))^{T},$ $(x_2, f(y_2))^{T}$ $ \in \cal C$.
Hence $\{x_1, y_1\}$, $\{x_1, y_2 \}$, $\{x_2, y_1\}$, $\{x_2, y_2\}$ are edges of $G$ forming a $4$-cycle, a contradiction.
So this case is impossible.

(2) There exist distinct elements $x_1, x_2, x_3 \in Q$ and distinct elements $f(y_1), f(y_2), f(y_3) \in Q$
such that $f(y_1)$, $f(y_2)$ $\in {\cal A}_{x_1}^1$, $f(y_2)$, $f(y_3)$ $\in {\cal A}_{x_2}^1$, $f(y_1), f(y_3) \in {\cal A}_{x_3}^1$.
In this case, $y_i$, $1 \leq i \leq 3$, are all distinct, and
$(x_1, f(y_1))^{T}$, $(x_1, f(y_2))^{T}$, $(x_2, f(y_2))^{T}$, $(x_2, f(y_3))^{T}$, $(x_3, f(y_3))^{T}$, $(x_3, f(y_1))^{T} \in \cal C$.
Hence $\{x_1, y_1\}$, $\{x_1, y_2\}$, $\{x_2, y_2\}$, $\{x_2, y_3\}$, $\{x_3, y_3\}$, $\{x_3$, $ y_1\}$
are edges of $G$ forming a $6$-cycle, a contradiction. So this case is not possible either.

Therefore, the $(2,M,q)$ code $\cal C$ constructed above is a $3$-MIPPC$(2,M,q)$ with $M =e(G)$.

This completes the proof. \end{IEEEproof}

Garc\'{i}a-V\'{a}zquez {\it et al.} \cite{GVBMV} stated that any maximum bipartite graph $G(q,q)$
with size $M(3,2,q)$ must have girth $8$, for $q\geq 6$ or $q=4$.
Therefore, we have the following corollary.

\begin{corollary}
\label{co.equiv222}
Let $q\geq 6$ or $q=4$. There exists a $3$-MIPPC$(2,M,q)$ if and only if
there exists a bipartite graph $G(q,q)$ of girth $8$ with $e(G)=M$.
\end{corollary}

\begin{lemma}{\rm (\cite{Neu})}
\label{Graph1111}
If $G(u,v)$ contains no cycle of length $4$ and $6$, then its size $e$ satisfies the following inequality
\[e^3 - (u + v)e^2 + 2uve - u^2v^2 \leq 0.\]
\end{lemma}

Then the size of a $3$-MIPPC$(2,M,q)$ can be derived from Lemmas \ref{equiv222} and \ref{Graph1111}.

\begin{corollary}
\label{bound1111}
For any $3$-MIPPC$(2,M,q)$, $M^3 - 2qM^2 + 2q^2M - q^4 \leq 0$.
\end{corollary}

Multimedia IPP codes are also closely related with generalized packings defined below.

\begin{definition}
\label{GP} Let $K$ be a subset of non-negative integers, and let $v,b$ be two positive integers.
A generalized $(v,b,K,1)$ packing is a set system $(X,{\cal B})$ where $X$ is a set of $v$ elements and
${\cal B}$ is a set of $b$ subsets of $X$ called blocks satisfying
\begin{itemize}
\item[(1)] $|B| \in K$ for any $B\in {\cal B}$;
\item[(2)] Every pair of distinct elements of $X$ occurs in at most one block of ${\cal B}$.
\end{itemize}
\end{definition}

A generalized packing $(X,{\cal B})$ is called $\triangle$-free if for any three distinct elements $P_1, P_2, P_3 \in X$,
if there are two blocks containing $P_1$, $P_2$ and $P_1$, $P_3$ respectively, then there is no block containing $P_2$, $P_3$.

\begin{theorem}
\label{equiv111}
There exists a $3$-MIPPC$(2,M,q)$ defined on $Q$ if and only if there exists
a $\triangle$-free generalized $(q,q,K,1)$ packing $(Q, \{{\cal A}_0^1, \ldots, {\cal A}_{q-1}^1\})$
with $K=\{|{\cal A}^1_0|, \ldots, |{\cal A}^1_{q-1}|\}$, and $M = |{\cal A}_0^1| + \cdots + |{\cal A}_{q-1}^1|$.
\end{theorem}
\begin{IEEEproof}  Suppose $\cal C$ is a $3$-MIPPC$(2,M,q)$ defined on $Q$,
and ${\cal A}_i^1 = \{b \in Q \  | \ (i, b)^T  \in \cal C\}$ for any $i \in Q$.
Then by Theorem \ref{cond111}, we know that $(Q, \{{\cal A}_0^1, \ldots, {\cal A}_{q-1}^1\})$
is a $\triangle$-free generalized $(q, q, \{|{\cal A}^1_0|, \ldots$, $|{\cal A}^1_{q-1}|\},1)$ packing,
and $M = |{\cal A}^1_0|+ \cdots + |{\cal A}^1_{q-1}|$.

Conversely, for any $\triangle$-free generalized $(q,q,K,1)$ packing $(Q, \cal B)$ with
${\cal B} = \{B_0, \ldots, B_{q-1}\}$ and $M = |B_0| + \cdots + |B_{q-1}|$, we define a set of vectors
${\cal B}^1 = \{B_0^1, \ldots, B_{q-1}^1\}$, with $B_i^1 = \{(i, b)^T \ | \ b \in B_i\}$ if $B_i \neq \emptyset$
and $B_i^1 = \emptyset$ if $B_i = \emptyset$, $0 \leq i \leq q-1$.
By Theorem \ref{cond111}, it is readily checked that ${\cal B}^1$ is a $3$-MIPPC$(2,M,q)$ defined on $Q$
and ${\cal A}_i^1 = B_i$ for any $i \in Q$.

This completes the proof. \end{IEEEproof}

\begin{corollary}
\label{co.equiv111}
There exists an optimal $3$-MIPPC$(2,M,q)$ if and only if there exists
a $\triangle$-free generalized $(q,q,K,1)$ packing with maximum $M = |{\cal A}_0^1| + \cdots + |{\cal A}_{q-1}^1|$,
where $K=\{|{\cal A}^1_0|, \ldots, |{\cal A}^1_{q-1}|\}$,
\end{corollary}

Now we show that some optimal $3$-MIPPC$(2,M,q)$s can be constructed by means of generalized quadrangles.

\begin{definition}
A finite generalized quadrangle (GQ) is an incidence structure ${\cal S} = (X, {\cal B}, I)$
with point-set $ X$ and line-set ${\cal B}$ satisfying the following conditions:
\begin{itemize}
\item[(1)] Each point is incident with $1+t$ lines ($t \geq 1$) and two distinct points are incident with at most one line;
\item[(2)] Each line is incident with $1+s$ points ($s \geq 1$) and two distinct lines are incident with at most one point;
\item[(3)] If $x$ is a point and $L$ is a line not incident with $x$, then
there is a unique pair $(y, N) \in X \times {\cal B}$ for which $xINIyIL$.
\end{itemize}
The integers $s$ and $t$ are the parameters of the GQ and ${\cal S}$ has order $(s, t)$; if $s = t$, ${\cal S}$ has order $s$.
\end{definition}

From the definition, any generalized quadrangle has no triangles.
It is known (see \cite{Pay}) that in a generalized quadrangle,
$|X| = (1+s)(1+st), |{\cal B}| = (1+ t)(1+st)$, and $s + t$ divides $st(1+s)(1+t)$.

\begin{lemma}
\label{constru1111}
If there exits a GQ$(s,t)$, then there exists a $\triangle$-free generalized $(v,b,1+s,1)$ packing,
where $v=(1+ s)(1+st), b= (1+t)(1+st)$.
\end{lemma}
\begin{IEEEproof} Suppose ${\cal S}=(X,{\cal B}, I)$ is a GQ$(s,t)$.
By regarding the lines of ${\cal S}$ as blocks and the points of ${\cal S}$ as elements,
we easily obtain a $\triangle$-free generalized $(v,b,1+s,1)$ packing $(X,{\cal B})$.
\end{IEEEproof}

\begin{lemma}{\rm (\cite{Pay})}
\label{GQresult}
Let $k$ be a prime power and $s \leq t$ be two positive integers.
Then there exist GQ$(s,t)$s for $(s,t) \in \{(k-1,k+1), (k,k), (k,k^2), (k^2,k^3)\}$.
\end{lemma}

If there exists a GQ$(s,t)$ with $s \le t$, then Lemma \ref{constru1111} gives
a $\triangle$-free generalized $(v,b,1+s,1)$ packing with $v=(1+s)(1+st) \le (1+t)(1+st) =b$.
Deleting $b-v$ blocks, we obtain a $\triangle$-free generalized $(v,v,1+s,1)$ packing.

\begin{corollary}
\label{mainres1111}
For any prime power $k$, there exist $3$-MIPPC$(2,M,q)$s for
$(M,q) \in \{(k^4,k^3), ((k^2+1)(k+1)^2,(k^2+1)(k+1)), ((k^3+1)(k+1)^2,(k^3+1)(k+1)), ((k^5+1)(k^2+1)^2,(k^5+1)(k^2+1))\}$.
\end{corollary}
\begin{IEEEproof} Apply Theorem \ref{equiv111} with Lemmas \ref{constru1111}, \ref{GQresult}.
\end{IEEEproof}

\begin{lemma}
\label{bound2222}
Let $a, d$ be two positive integers with $d^2 - 2d + 2 - a = 0$. Then for any $3$-MIPPC$(2,M,ad)$, we have $M \leq ad^2$.
\end{lemma}
\begin{IEEEproof} For any $3$-MIPPC$(2,M,q)$, by Corollary \ref{bound1111},
we know that $M^3 - 2qM^2 + 2q^2M - q^4 \leq 0$.
Let $f(M) = M^3 - 2qM^2 + 2q^2M - q^4$, then the derivative of $f(M)$ is
\[\frac{df}{dM}(M) = 3M^2 - 4qM + 2q^2 = 3(M -\frac{2q}{3})^2 +\frac{2q^2}{3} > 0. \]
Therefore, $f$ is a strictly increasing function on $M$.
Let $q = ad$, where $a$ and $d$ are positive integers such that $d^2 - 2d + 2 -a = 0$. Then
\begin{eqnarray*}
\begin{split}
f(ad^2) & =  (ad^2)^3 - 2(ad)(ad^2)^2 + 2(ad)^2(ad^2) - (ad)^4 \\
& =  a^3d^6 - 2a^3d^5 + 2a^3d^4 -a^4d^4 \\
& =  a^3d^4(d^2 - 2d + 2 -a) \\
& =  0.\end{split}
\end{eqnarray*}
For any $M^{'} > ad^2$, we have $f(M^{'}) > 0$.
So $ad^2$ is the greatest integer which satisfies the inequality $M^3 - 2qM^2 + 2q^2M - q^4 \leq 0$.
This completes the proof.\end{IEEEproof}

\begin{theorem}
\label{optimal}
There exists an optimal $3$-MIPPC$(2, (k^2+1)(k+1)^2, (k^2+1)(k+1))$ for any prime power $k$.
\end{theorem}
\begin{IEEEproof} A $3$-MIPPC$(2, (k^2+1)(k+1)^2, (k^2+1)(k+1))$ exists from Lemma \ref{mainres1111}.
Let $a = k^2+1, d = k+1$, then $d^2 - 2d + 2 -a = 0$. Apply Lemma \ref{bound2222}.
\end{IEEEproof}

\section{Asymptotically Optimal $3$-MIPPC$(2,M,q)$s}

Corollaries \ref{co.equiv222} and \ref{co.equiv111} inspire us to construct optimal $3$-MIPPC$(2,M,q)s$
via bipartite graphs with girth $8$ or maximum $\triangle$-free generalized $(q,q,K,1)$ packings.
Unfortunately, except for the result in Theorem \ref{optimal},
we do not know other infinite families of optimal $3$-MIPPC$(2,M,q)$s.
However, we can construct several infinite families of asymptotically optimal $3$-MIPPC$(2,M,q)$s
by truncating points and lines from generalized quadrangles.

\begin{theorem}
\label{asymopti111}
There exists a $3$-MIPPC$(2, k^4 + 2k^3 + 2k^2 + 2k -2sk, k^3 + k^2 + k + 1 - s)$
for every prime power $k$, where $1 \leq s \leq k^2 + k + 1$.
\end{theorem}
\begin{IEEEproof} If we can construct a $\triangle$-free generalized $(k^3+k^2+k+1-s, k^3+k^2+k+1-s, \{k,k+1\},1)$ packing
with $k^3+k^2+k-sk$ blocks of size $k+1$ and $sk-s+1$ blocks of size $k$,
then the conclusion would follow from Theorem \ref{equiv111}.
According to Lemma \ref{GQresult}, there exists a GQ$(k,k)$, say ${\cal S}=(X,{\cal B},I)$, for every prime power $k$.
Choose an arbitrary point $x_{0,0} \in X$.
Let $L_{0,j} = \{x_{0,0}, x_{1,j}, \ldots, x_{k,j}\}$, $0 \leq j \leq k$, be the $k+1$ distinct lines incident with $x_{0,0}$,
and $L_{i,1}, \ldots, L_{i,k}$, $1 \leq i \leq k$, be the other $k$ distinct lines incident with $x_{i,0} \in X$.
Let $s_1 = \lfloor\frac{s-1}{k}\rfloor$ and $s_2 = s-1- ks_1$.
Then the desired $\triangle$-free generalized packing can be constructed by eliminating
$s$ points $x_{0,0}$, $x_{1,0}$, $\ldots$, $x_{k,0}$, $x_{1,1}$, $\ldots$, $x_{k,1}$, $\ldots$, $x_{1,s_1-1}$, $\ldots$, $x_{k,s_1-1}$, $x_{1,s_1}$, $\ldots$, $x_{s_2,s_1}$ and
$s$ lines $L_{0,0},L_{0,1},\ldots,L_{0,k},$ $L_{1,1}$, $\ldots$, $L_{1,k}$, $\ldots$, $L_{s_1-1,1}$, $\ldots$, $L_{s_1-1,k}$, $L_{s_1,1}$, $\ldots$, $L_{s_1,s_2}$,
where the size of each line after elimination is $k+1$ or $k$ because of the ${\triangle}$-freeness of the GQ.
\end{IEEEproof}

\begin{theorem}
\label{asymopti222}
There exists a $3$-MIPPC$(2,k^4-sk,k^3-s)$ for every prime power $k$, where $0 \leq s \leq 2k - 1$.
\end{theorem}
\begin{IEEEproof} Similar to Theorem \ref{asymopti111}, we want to construct a $\triangle$-free generalized $(k^3-s,k^3-s,\{k\},1)$ packing.
According to Lemma \ref{GQresult}, there exists a GQ$(k-1,k+1)$, say ${\cal S}=(X,{\cal B},I)$, for any prime power $k$.
Then $|X| = k^3$ and $|{\cal B}| = k^3 + 2k^2$.
Let $x_0 \in X$ and $X_0 = \{x \in X\setminus\{x_0\} \ | \ x_0 \ \mbox{ and} \ x  \mbox{ are  incident with  a  line}\}$.
Then $| X_0 | = k^2 + k -2$.
Let $X_s = \{x_0, x_1, \ldots, x_{s-1}\} \subseteq \{x_0\} \cup X_0$ and
${\cal B}_s = \{L \in {\cal B} \ | \ L \mbox{ is incident with a point}  \ x \\ \in X_s\}$.
By a simple counting argument, we know that $|{\cal B}_s| = (k+2)+(s-1)(k+1) = s + sk + 1$.
Then we can obtain a $\triangle$-free generalized $(v,b,k,1)$ packing by eliminating the $s$ points in $X_s$
and the $s+sk+1$ lines in ${\cal B}_s$ from the GQ$(k-1,k+1)$, ${\cal S}$, where $v=k^3-s$ and $b=k^3-s+(2k^2-sk-1)$.
Since $0 \leq s \leq 2k-1$, we have $b \ge v$.
Therefore the desired $\triangle$-free generalized packing exists by further eliminating $b-v$ blocks of
the $\triangle$-free generalized $(v,b,k,1)$ packing.
\end{IEEEproof}

\begin{theorem}
\label{asymopti333}
There exists a $3$-MIPPC$(2, k^4+2k^3+2k^2-sk-s+\lfloor\frac{s-1}{k+1}\rfloor, k^3+2k^2-s)$
for every prime power $k$, where $1 \leq s \leq k^2+k+1$.
\end{theorem}
\begin{IEEEproof}
According to Lemma \ref{GQresult} and the point-line duality of GQs (see, for example, \cite{Pay}),
there exists a GQ$(k+1,k-1)$ for any prime power $k$.
Suppose that ${\cal S}$ is a GQ$(k+1, k-1)$. Then $|X| = k^3+2k^2$ and $|{\cal B}| = k^3$. Pick an arbitrary point $x \in X$.
Suppose $L_{i}=\{x,x_{i,1},\ldots,x_{i,k+1}\}$,  $1 \leq i \leq k$, are $k$ distinct lines containing $x$,
and each $P_{i}$ is the point-set of $L_{i}$.
Let $s_1= \lfloor\frac{s-1}{k+1}\rfloor$, $s_2 = s-1 -s_1(k+1)$, and
\[{\cal P}_s = \left\{\begin{array}{l}
\{x\}, \ \ \ \ \ \ \ \ \ \ \ \
  {\rm if} \ \  s=1,\\[2pt]
\{x\}\bigcup (\bigcup\limits_{i=1}^{s_1}{P_i}),
 {\rm if} \  s \neq 1 \  {\rm and} \  s\equiv 1 \pmod {k+1},\\[2pt]
\{x\}\bigcup (\bigcup\limits_{i=1}^{s_1}{P_i}) \bigcup \{x_{s_1+1, 1}, \cdots, x_{s_1+1, s_2}\},  {\rm otherwise}.
\end{array}
\right.
\]

For a given $s$, we can eliminate the point-set ${\cal P}_s$ and derive a $\triangle$-free generalized
$(v,b,\{k+1-s_2, k+1, k+2\},1)$ packing with $(s-1)(k-1)+k-s_1- h(s_2)$ blocks of size $k+1$,
$k^3-k-(s-1)(k-1)$ blocks of size $k+2$, and $h(s_2)$ block of size $k+1-s_2$,
where $v = k^3+2k^2-s$, $b = k^3-s_1$, and
\[h(s_2) = \left\{\begin{array}{rl}
0,  & {\rm if} \ s_2=0,\\[2pt]
1,  & {\rm otherwise.}
\end{array}
\right.
\]
Then $v-b = 2k^2-s+s_1 >0$. So, the desired generalized packing can be constructed by adding $v-b$ blocks
containing exactly one point belonging to $X \setminus {\cal P}_s$. Now we compute the value $M$.
\begin{eqnarray*}
M = [(s-1)(k-1)+k-s_1- h(s_2)](k + 1)\ \ \ \ \ \ \ \ \ \ \ \ \ \ \ \ \ \ \ \ \ \ \ \ \ \ \ \ \ \ \ \ \ \ \ \ \\
+ [k^3-k - (s-1)(k-1)](k+2) \ \ \ \ \ \ \ \ \ \ \ \ \ \ \ \ \ \ \ \ \ \ \ \ \ \ \ \ \ \ \ \ \ \ \ \ \ \ \ \ \ \ \ \\
+ h(s_2)(k+1-s_2) + 2k^2-s+s_1\ \ \ \ \ \ \ \ \ \ \ \ \ \ \ \ \ \ \ \ \ \ \ \ \ \ \ \ \ \ \ \ \ \ \ \ \ \ \ \ \ \ \\
= k^4 +2k^3 + 2k^2 - sk- s_1k -1 - h(s_2)s_2.\ \ \ \ \ \ \ \ \ \ \ \ \ \ \ \ \ \ \ \ \ \ \ \ \ \ \ \ \ \ \ \ \ \
\end{eqnarray*}
If $s_2 \neq 0$, then $h(s_2)s_2 = s_2$;
if $s_2 =0$, then $h(s_2)s_2 = 0 =s_2$.
So\\
$M = k^4 +2k^3 + 2k^2 -sk-s_1k -1 - s_2\\
\indent = k^4 +2k^3 + 2k^2 -sk-s_1k -1 - (s - 1 - s_1(k+1))\\
\indent = k^4 +2k^3 + 2k^2 -sk -s - s_1\\
\indent = k^4 +2k^3 + 2k^2 -sk -s - \lfloor \frac{s-1}{k + 1}\rfloor$.

This completes the proof.\end{IEEEproof}

\begin{theorem}
\label{asymopti444}
The $3$-MIPPC$(2,M,q)$s constructed in
Theorems {\rm \ref{asymopti111}, \ref{asymopti222} and \ref{asymopti333}} are asymptotically optimal.
\end{theorem}
\begin{IEEEproof} Here, we only prove that the $3$-MIPPC$(2,M,q)$s constructed in
Theorem \ref{asymopti222} are asymptotically optimal. The other two cases can be proved in a similar way.
Note that in Theorem \ref{asymopti222}, $q = k^3-s$, $M = k^4-sk$, where $k$ is a prime power and $0 \leq s \leq 2k-1$.

Just as in the proof of Lemma \ref{bound2222},
we consider the strictly increasing function $f(M) = M^3 - 2qM^2 + 2q^2M - q^4$, and also the cubic equation $f(M) = 0$.
Let $a=1, b=-2q, c=2q^2, d=-q^4$.
Then the discriminant of the above-mentioned cubic equation is
$D=18abcd-4b^3d+b^2c^2-4ac^3-27a^2d^2=q^6(40q-16-27q^2)<0$,
which implies that this cubic equation has one real root $M_0$ and two complex conjugate roots
(see, for example, \cite{Ir}, and also \cite{Neu}), where\begin{small}
\begin{eqnarray*}
\begin{split}
M_0  = &\ -\frac{b}{3a}  -\frac{1}{3a}\sqrt[3]{\frac{1}{2}[2b^3-9abc+27a^2d+\sqrt{-27a^2D}]} \\[0.2cm]
            & \    -\frac{1}{3a}\sqrt[3]{\frac{1}{2}[2b^3-9abc+27a^2d-\sqrt{-27a^2D}]} \\[0.2cm]
          = & \ \frac{2q}{3}  -\frac{q}{3}\sqrt[3]{\frac{1}{2}[20-27q+\sqrt{27(27q^2-40q+16)}]} \\[0.2cm]
            & \   -\frac{q}{3}\sqrt[3]{\frac{1}{2}[20-27q-\sqrt{27(27q^2-40q+16)}]}.
\end{split}\end{eqnarray*}\end{small}
Noting that $f(0) = -q^4 <0$, we have $M_0 > 0$.
By Corollary \ref{bound1111},  $M(3,2,q) \le M_0$, and then
$0 < \frac{M}{M_0} \le \frac{M}{M(3,2,q)}$ $ \le 1$.
Therefore it is sufficient to prove that $\lim\limits_{q \rightarrow \infty} \frac{M}{M_0} = 1$ holds.

Since $q=k^3-s$, we have\begin{small}
\begin{eqnarray*}
\begin{split}
&\lim\limits_{q \rightarrow \infty} \frac{M_0}{k^4} =  \lim\limits_{k \rightarrow \infty} \frac{M_0}{k^4} \\
        =& \lim\limits_{k \rightarrow \infty} \frac{2q}{3k^4}    - \lim\limits_{k \rightarrow \infty} \frac{q}{3k^4}\sqrt[3]{\frac{1}{2}[20-27q+\sqrt{27(27q^2-40q+16)}]} \\[0.2cm]
       &    - \lim\limits_{k \rightarrow \infty} \frac{q}{3k^4}\sqrt[3]{\frac{1}{2}[20-27q-\sqrt{27(27q^2-40q+16)}]}   \\
        = & \ 0-0-(-1) \\
        = & \ 1,
       \end{split}
\end{eqnarray*}\end{small}
then
\begin{eqnarray*}
\lim\limits_{q \rightarrow \infty} \frac{M}{M_0} = \lim\limits_{k \rightarrow \infty} \frac{M}{M_0}  = \frac{\lim\limits_{k \rightarrow \infty} \frac{M}{k^4}}{\lim\limits_{k \rightarrow \infty} \frac{M_0}{k^4}}
         =  \frac{1}{1}
          =  1.
\end{eqnarray*}
This completes the proof.\end{IEEEproof}

\section{Concluding Remarks}

In this paper, we introduced multimedia IPP codes, which can be used to identify
at least one malicious authorized user in a multimedia fingerprinting system.
We characterized an optimal $3$-MIPP code of length $2$ in terms of a maximum bipartite graph with girth $8$
and a $\Delta$-free generalized packing with maximum number of points in all blocks, respectively.
By using bipartite graphs, we derived several upper bounds on the size of a multimedia IPP code.
By using $\Delta$-free generalized packings, we constructed several infinite families of
(asymptotically) optimal $3$-MIPP codes of length $2$ via generalized quadrangles,
which can be used to construct ``good" binary $3$-MIPP codes with long length by a simple composition construction,
in the sense that all these codes have quite a few codewords.

It would be interesting if we could find more optimal multimedia $t$-IPP codes.
However, we do not find it easy to construct optimal multimedia $t$-IPP codes with long length $n$,
even for $n=4$.

\section{Acknowledgments} %

 Cheng, Jiang and Miao thank Professor Gennian Ge for his helpful discussions on
$\Delta$-free generalized packings, bipartite graph with high girth, and generalized quadrangles.





\end{document}